\documentclass{elsarticle}
\usepackage{epsfig,graphicx}
\usepackage[latin1]{inputenc}
\usepackage[english]{babel}
\usepackage{amsmath}
\usepackage{amsfonts}
\usepackage{amssymb}
\usepackage{graphicx}
\usepackage[left=2cm,right=2cm,top=2cm,bottom=2cm]{geometry}

\usepackage{color}
\usepackage{float}
\usepackage{afterpage}
\usepackage{xcolor}
\colorlet{darkred}{red!80!black}
\colorlet{darkgreen}{green!50!black}
\usepackage[normalem]{ulem}

\def\lp {\left( }
\def\rp {\right) }
\def\lb {\left[ }
\def\rb {\right] }

\def\bea{\begin{eqnarray}}
\def\eea{\end{eqnarray}}
\def\nn {\nonumber}

\def\D {\Delta}

\def\p {\pi}

\newcommand{\bkkk}{B^+ \to  K^-K^+K^+}

\newcommand{\bckkp}{B^+_c \to  K^-K^+\pi^+}

\begin{document}
 \makeatletter
\begin{frontmatter}
\title{ Charm rescattering contribution in rare $B^+_c\to K^+K^-\p^+$ decay}
\author[CBPF]{I. Bediaga} 
\author[ITA]{T. Frederico}
\author[CBPF,TUM]{P. C. Magalh\~aes}
\ead{patricia.magalhaes@tum.de}
\address[CBPF]{Centro Brasileiro de Pesquisas F\'isicas, \\
22.290-180, Rio de Janeiro, RJ, Brazil}
\address[ITA]{Instituto Tecnol\'ogico de
Aeron\'autica, DCTA \\ 12.228-900 S\~ao Jos\'e dos Campos, SP,
Brazil.}
\address[TUM]{Department of Physics, Technische Universität München,\\ 85748 Garching, Germany.}

\begin{abstract}
Following the experimental results from LHCb on the rare decay $\bckkp$, 
we investigate the possibility where this process is dominated by a  double charm rescattering. 
The $B_c$  decay to double charm channels have a weak topology that is favoured in comparison with the direct production of $K^-K^+\pi^+$ in the final state, suppressed by quark annihilation.
The decay amplitude for  $\bckkp$ with $B_c$ decaying first to double charm channels is described 
  by charm hadronic  triangle loops, which reach the final state of interest 
after $D\bar{D}\to K\bar{K}$  or $ D^+ D^-_s \to \pi^+  K^-$  transitions.
We show that these processes give rise to non-resonant amplitudes with a clear signature in the Dalitz plot. In a near future, 
the new data from LHCb run II will be able to confirme if the main hypotheses of this work
 is correct and the dominant mechanism to produce $K^+K^+\pi^-$ from the decay of $B^+_c$ is through charm rescattering. 
\end{abstract}
\end{frontmatter}

\section{Introduction}
Recently a LHCb experiment~\cite{adlene} reported  the data for  the $\bckkp$ decay. Evidence was presented for a 4$\sigma$ signal for the $B_c^+ \to \chi_{c0}(\to K^+K^-)\pi^+$ and an indication of 2.4 $\sigma$ 
signal for the $\bckkp$  decay out of  $\chi_{c0}$ mass region, using  run I data. 
This rare decay at quark tree level is produced only by the suppressed  W-annihilation topology.
However,  one can  also consider the possibility that this $K^+K^-\pi^+$ final state is produced by a favoured topology, 
 like the allowed double charm $B_c$
 decay channel,  followed by  the rescattering  of the charm mesons  leading to  light pseudoscalar channels. 
 The importance of rescattering in nonleptonic B  decays has been discussed in different theoretical 
approaches \cite{us, BOT, ABAOT, GLR, Soni2005, patig, CPV}. 
In this context the $\bckkp$ channel, suppressed at quark level and favoured by double charm final states, is a good candidate to study the importance of hadronic final state interactions (FSI) to form the observed decay channels.  
In this particular decay the rescattering mechanism is proposed to be driven by a triangle hadronic loop  \cite{us, Soni2005}.

In a recent work we have also investigated the  charm penguin contribution in $\bkkk$ decay \cite{us} for two different regions of the phase
 space:  the nonperturbative region, described by a hadronic triangle loop, and the quasi perturbative region described by a 
quark loop. The former includes a $D\bar{D}\to K\bar{K}$  transition  for which we proposed a  scattering amplitude inspired by the 
Regge expansion,  considering the damping factor  of the S-matrix with off-shell effects. 
 In our studies for $\bkkk$ decay \cite{us} we showed that both  penguin loops, i.e. the double charm quark level loop 
and the hadronic triangle with charm mesons propagators,  will present a particular signature in Dalitz plot. 
 
In another theoretical approach  Gronau, London and Rosner~\cite{GLR}  proposed a method to estimate  the 
relative importance of the double charm rescattering to light mesons in nonleptonic B  decays  for weak processes 
that at the fundamental level generally occur via the suppressed  quark annihilation and W exchange diagrams.
They size the amplitude of the diagram associated with the dominant topology  based on 
known branching fractions of processes with similar topologies.  
The authors predicted branching fractions, including the $B^0 \to K^+K^-$, also suppressed at quark level by W-exchange 
(still unseen at that time), which  agrees with recent LHCb results~\cite{LHCb_bkk}. 
Other important issue raised by reference \cite{GLR} was  the quantification of hadronic rescattering to different channels
including the double charm to light pseudoscalar mesons, such as $B_s\to D^+_s D_s^- \to \p^+\p^-$. 

   In this work, we conjecture that the $\bckkp$ decay occurs  mainly 
through  the production of Cabbibo-favoured double charm meson  states
%states  as the} charm penguin processes\footnote{{\color{red} Note that the reference to charm penguin is based on Ref. \cite{us} and thing of quark penguin diagram.} }
  followed  by the   hadronic rescattering to the $ K^+K^-\p^+$ channel. 
We propose to investigate the Dalitz plot signatures of 
these contributions by performing an analysis following closely our previous work \cite{us}.  
In the near future we can expect that the new results from LHCb run II increase the
 statistics presenting  a better definition of the branching fraction and event distribution
 in the Dalitz plot for  $\bckkp$, which can confirm or disprove the dominance of  the double-charm rescattering 
 to form  this particular decay channel.

\section{Rescattering contribution to rare $B$ decay }

Hadronic decays of the $B_c$ meson, the heaviest one established so far, are 
particularly intriguing. The dominant decay channels are composed of two heavy mesons in the final state, 
while decays directly producing  light hadrons in final states at the fundamental level are  suppressed  by 
the weak annihilation topology of the quark  processes.
Another important characteristic of $B_c$ decays to light mesons is the huge phase space 
available, which allows contributions from  charm and bottom mesons  as intermediate states, 
as well as  many rescattering combinations from double charm  or exotics.  
   
In the case of $\bckkp$ one can see in Fig. 1 different possibilities of 
  producing $KK\p$ in the final state: a) directly from a tree level diagram, 
and b) and c) double charm production hadronizing to a pair of charmed mesons, which radiates a light meson and
rescatters to a pair of light mesons, creating the final three-body decay channel via a hadronic process,
 as illustrated in Fig. \ref{fig2}.
All diagrams are proportional to the same CKM elements, but in one side the
 direct process is known to be strongly suppressed by helicity conservation, whereas the 
rescattering contributions introduce a damping factor from the loop itself, as one could see from Eq. (\ref{A1}).

\begin{figure}[hbt]
\begin{center}
\includegraphics[width=.5\columnwidth,angle=0]{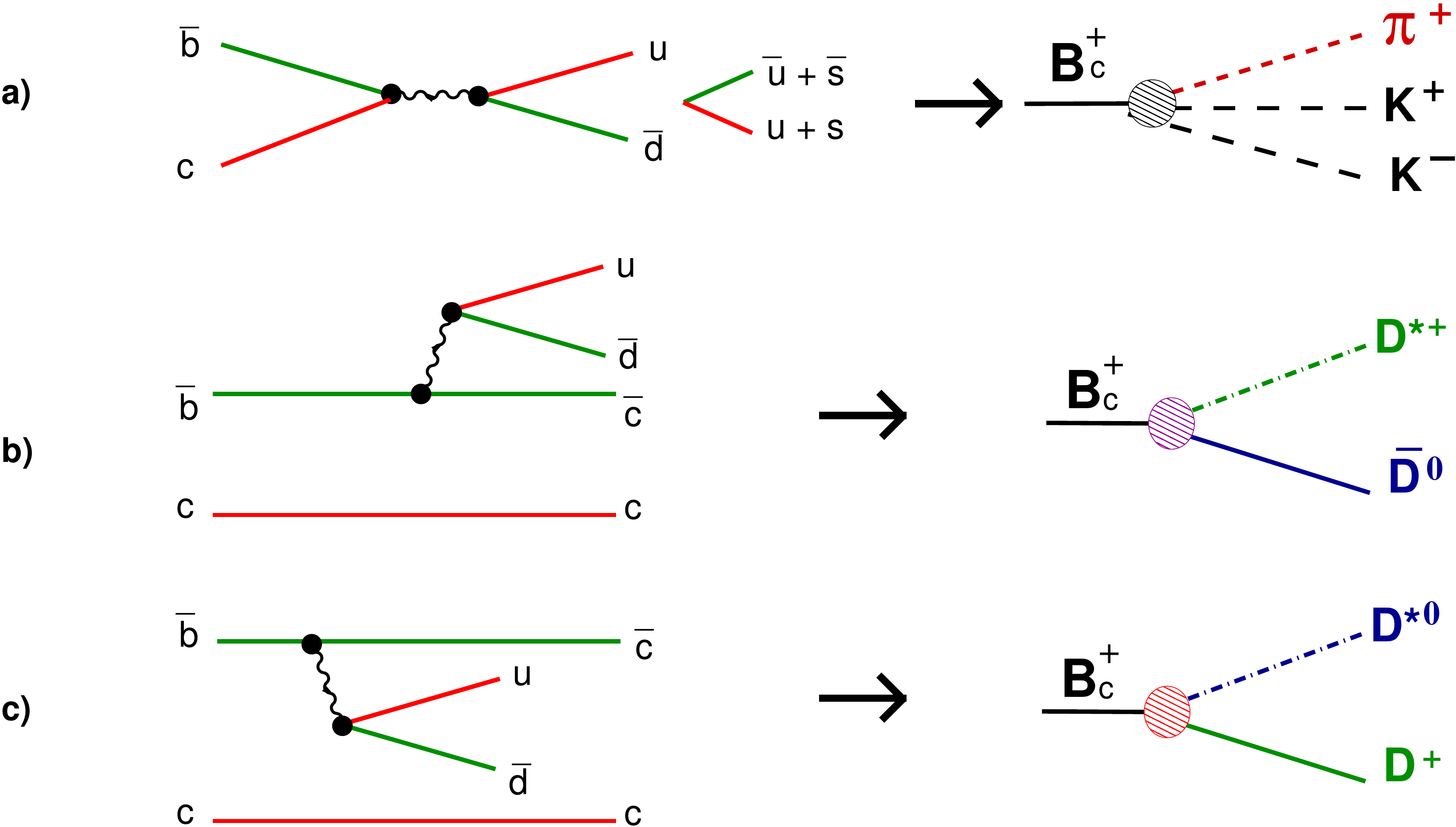}
{\small
 \caption{ Quark level  diagrams for the different topologies 
contributing to $B^+_c \to K^+K^-\pi^+ $ decay and to double charm production: (a) direct tree decay; 
(b) $ B^+_c \to {D^ {*+}} \bar D^ 0 $  (c.f. Fig. \ref{fig2} left diagram);
(c)  $ B^+_c \to D^{*0} D^+$  (c.f. Fig. \ref{fig2} right diagram).}}
\end{center}
\label{fig1}
\end{figure}

As discussed by  Gronau, London and Rosner \cite{GLR}, the hadronic  rescattering 
from topologies that are favoured, even when considering the suppression introduced by 
the rescattering amplitude,  can give a significant contribution to the non-favoured  light
meson three-body decay channel, which might leave  a signature in the data.  
In order to identify this signature we first have to build the three-body decay amplitude,
starting from the favoured charmed meson two-body decay channels.

 There are a number of double charm meson intermediate states that can contribute to the $K^-K^+\pi^-$ 
final state of the $B_c$ decay. Following reference \cite{GLR} we select only  processes
 with large branching fractions to be  considered as the dominant ones. %s to form this particular light-meson three-body  final state.
 In the $K\bar{K}$ channel we consider  the $\bckkp$ decay to be mediated by 
$D^{*+}(2010) \bar{D^0}$, represented by the hadronic 
triangle loop shown in Fig. \ref{fig2} (left).  
The branching fraction for $B_c^+ \to D^{*+}(2010) \bar{D^0}$ 
was measured to be of the order of $10^{-3}$ \cite{pdg}; whereas  $ D^{*+}(2010) \to D^0\pi^+ $
 accounts for $\approx 67\%$ of  $D^{*+}(2010)$ width \cite{pdg}. A similar double charm 
rescattering contribution is expected in the $K^-\pi^+$ channel,  where
  $B_c^+ \to D^{*0} \bar{D^+} $ is followed by the virtual decay  $D^{*0} \to D^-_s K^+$ and the rescattering  
 $D^+ D_s^- \to \pi^+ K^-$ as illustrated in Fig.~\ref{fig2} (right).
 Although the relative branching ratios are unknown, we expect them to be at same order of the 
previous one.

To fully calculate the two hadronic heavy loops presented in Fig.~\ref{fig2},
 we need the rescattering amplitudes 
$D^+ D_s^- \to \pi^+ K^-$ and $\bar{D^0} D^0 \to K^+K^-$. The latter was 
developed in detail in our previous study of $\bkkk$ decay \cite{us} and can be adapted to
describe the first one. 

\begin{figure}[ht]
\begin{center}
\includegraphics[width=.35\columnwidth,angle=0]{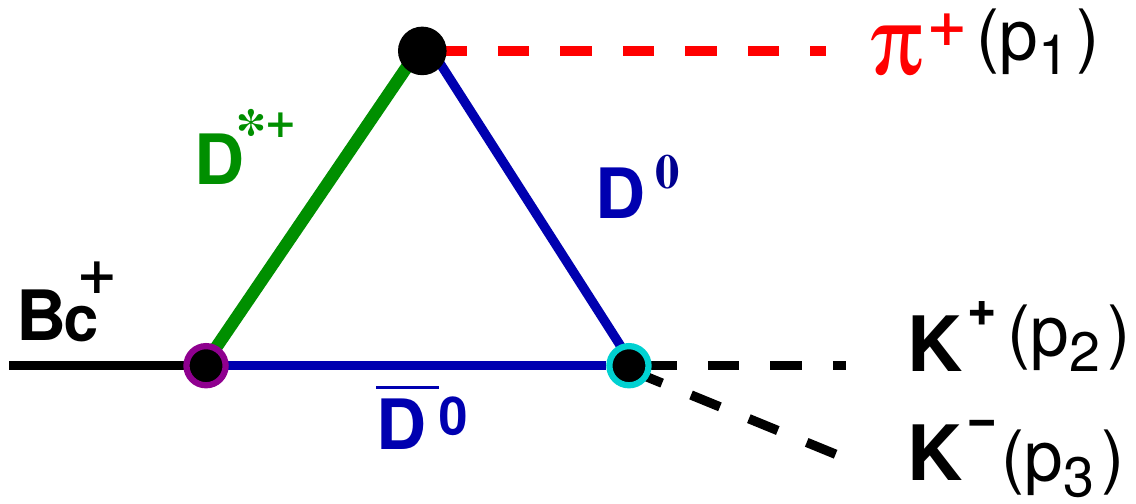}
\hspace*{1cm}
\includegraphics[width=.35\columnwidth,angle=0]{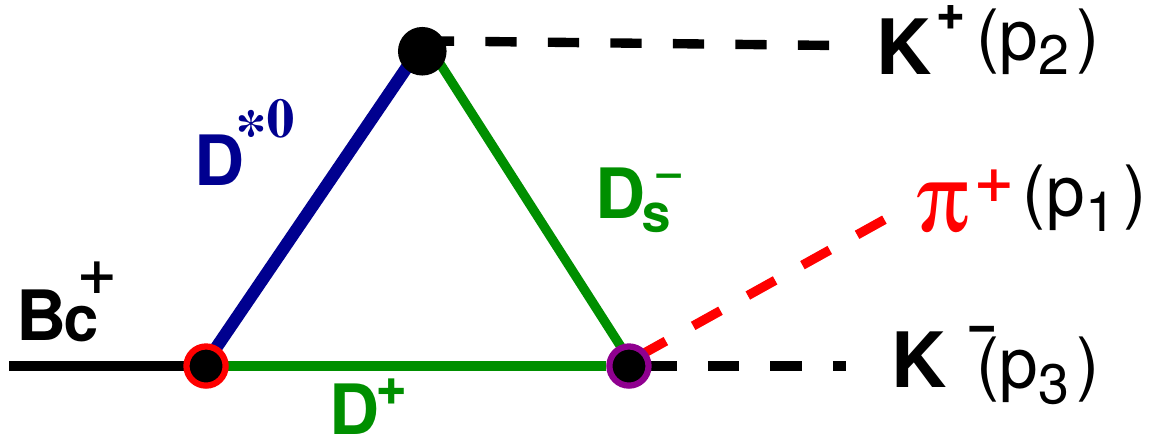}
\caption{Charm hadronic loop contribution to $\bckkp$ decay. The final state $K^-K^+\p^+$ is reached 
after $\bar{D^0} D^0 \to K^+K^-$ or $D^{+} D^-_s \to \pi^+K^-$ rescattering.  Left diagram: 
pion emission  by the $D^{*+}$ followed  by  $ \bar{D^ 0} D^ 0 \to K^+  K^-$ rescattering. Right diagram: 
kaon emission by the $D^{*0}$ followed  by   $ D^+ D^-_s \to \pi^+  K^-$ rescattering. }
\end{center}
\label{fig2}
\end{figure}

\section{ Hadronic loop }

Hadronic triangle loops were investigated previously in a number of processes in 
$D$ and $B$ three-meson decays
~\cite{us, patig, patriciaPRD, karinJHEP, patWeak}. 
 The presence of these particular modes of three-body final state interaction 
 were reported  to play an important role also in kinematical regions of the phase-space different from the one where
 the momentum  is shared between all the particles in the final state.
 In the case of $\bckkp$, given the suppression of the  process represented by the diagram a) of Fig. 1, 
  we assume that there is no tree level contribution from the meson formation directly after the partonic process.  
 Then, the decay amplitude is assumed to be described mainly by the 
 loop diagrams in Fig. \ref{fig2}.
 These triangle loops are very similar but contribute to different 
channels: $KK$ and $\pi K$. Therefore, in what follows we detail only the calculation which includes the double charm rescattering 
to the $KK$ channel.

Following the formalism developed in \cite{us, patig, patriciaPRD, patWeak} we obtain the amplitude for
 $\bckkp$ decay  from the left diagram in Fig. 2  as a function of the $K^+K^-$ invariant mass, $ m^2_{KK} = s_{23}$,
 which can be written as:
 \bea
A^{KK}_{B_c} (s_{23})= i  \int \frac{d^4 \ell}{(2\p)^4} \; 
\frac{T_{B_c\to D^{*+}\bar{D^0}}(l^2)\,T_{\bar{D^0}D^0\to KK}(s_{23})}{\D_{D^{+*}}\D_{D^0}
 \,\D_{\bar{D^0}} }\;,
\label{A1}
\eea
where   $\left[\D_{D^i}\right]^{-1}$ are the charm propagators inside 
the loop:
\bea
\D_{D^0} = (l-p_3)^2 - M^2_{D^0}\,, \,\,\,\,\,\,\,\,\D_{\bar{D^0}} = (l-P_B)^2 - M^2_{D^0} 
\,, \,\,\,\,\, \D_{D^{+*}} =l^2 - \Theta_{D*+} \,, \nn \,\,\,\, 
\eea
 with $\Theta_{D*+} = M_{D^{*+}}^2 - i\Gamma_{D^{*+}}/2$.
The amplitude  $T_{B_c\to D^{*+}\bar{D^0} } (l^2)$ accounts for the weak couplings: $B_c\to
 W\bar{D^0} $ and $W\to D^{*+}$,  with the former described by form factors with 
a single pole approximation with a nonzero width:
\bea
F^+_{B_c\to\bar{D^0}} (l^2) &=& F_0 \frac{ - m^2_a}{\D_a(l^2)}, \,\,\,
 \D_a(l^2)= l^2 - (m^2_a -i\,\Gamma_a) .
\eea
$T_{\bar{D^0}D^0\to KK}$ is the rescattering amplitude for $\bar{D^0}D^0\to K^+K^-$ described by a 
phenomenological model taking into account the S-matrix unitarity \cite{us} (for details see Appendix 2 of
 Ref.~\cite{us}). The model also takes into account the off-shell contribution below the double charm threshold $s_{th\,D\bar{D}}$:
\bea 
T_{\bar{D^0}D^0\to KK}(s) &= & \frac{s^\alpha}{s_{th\,D\bar{D}}^\alpha}\, \frac{2\kappa_2}{\sqrt{s_{th\,D\bar{D}}}}\,\left({s_{th\,D\bar{D}}\over s +s_{QCD}}\right)^{\xi+\alpha}
\left[\left({ c + bk_1^2-i k_1 \over c + bk_1^2 +i k_1}\right)\,\,
\left({\frac{1}{a}+ \kappa_2\over\frac{1}{a}-\kappa_2}\right)\right]^\frac12, \,\,s < s_{th\,D\bar{D}}
\label{t12c} 
\\[2mm]
&=&- i\, \,\frac{2\,k_2}{\sqrt{s_{th\,D\bar{D}}}}\left({s_{th\,D\bar{D}}\over s +s_{QCD}}\right)^{\xi}
\left({m_0\over s - m_0}\right)^{\beta} 
\left[\left({ c + bk_1^2-i k_1 \over c + bk_1^2 +i k_1}\right)
\left({\frac{1}{a}-i k_2\over\frac{1}{a}+i k_2}\right)\right]^\frac12, s\geq s_{th\,D\bar{D}}
\nn\label{t12d} 
\eea
 where  $k_1=\frac12\sqrt{s-s_{th\,K\bar{K}}}$ and  $k_2=\frac12\sqrt{{s-s_{th\,D\bar{D}}}}$ are the center mass momentum  above the threshold and 
$\kappa_1=\frac12 \sqrt{{s_{th\,K\bar{K}} -s}}$ 
is the rest frame double charm momentum below the threshold.
The model parameters are given in Table \ref{tab:1}.
\begin{table}
\caption{Parameters used in  the phenomenological scattering amplitude given by Eq.(\ref{t12c})~\cite{us}. 
The values for the $D^{-}D_s^{+} \to K^-\pi^+$ scattering amplitude are the same as 
$D^{0}\bar{D^{0}} \to K^+K^- $ except for the low and high energy thresholds. }
\begin{center}
\begin{tabular}{|c|c|c|c|c|c|c|c|c|c|c|c|}
\hline
$\alpha$& $\xi$& c& b& a& $s_{QCD}$&$m_0$& $\beta$ &$s_{th\,K\bar{K}}$ & $s_{th\,D\bar{D}}$ &$s_{th\,K\pi}$ & $s_{th\,DD_{s}}$\\ 
\hline 3 & 2.5 &  0.2 & 1 &  -$\infty$ & 0.2 GeV$^2$ &  8  &2 
& 0.97 GeV$^2$ & 13.85 GeV$^2$ & 0.40 GeV$^2$ & 14.72 GeV$^2$\\
\hline 
\end{tabular}
\end{center}
\label{tab:1}
\end{table}

The $\bckkp$ decay amplitude is then given by:
\bea
A^{KK}_{B_c}(s_{23}) = i C\,\, \,m^2_a \int \frac{d^4 \ell}{(2\p)^4} \; 
\frac{T_{\bar{D^0}D^0\to KK}(s_{23}) \,[-2\,p'_3\,\cdot(p'_2-p_1)] }{\, \D_{D^{+*}}\D_{D^0}
 \,\D_{\bar{D^0}}\,\D_a}\;,
\label{A1.1}
\eea
where $C$ includes the  weak vertex and unknown coupling constants, and 
 $p'_i$ is the momentum of a given $D$ meson inside the loop. 
  The scalar product in Eq.(\ref{A1.1}) can be written in terms of the meson propagators
 and the loop integral momentum $l$ as
\bea
-2\,p'_3\,\cdot(p'_2-p_1) &=& \,[\, \D_{D_0} +\,\D_{\bar{D_0}} - \,2\,s_{23} + \,M^2_\p+
 2\,M^2_{D_0}  - l^2 \,]\
\eea
and finally 
\bea
A^{KK}_{B_c}(s_{23}) = i \,C \,m^{2}_{a} \,\int \frac{d^4 \ell}{(2\p)^4} \; 
\lb \,T_{DD\to KK}(s_{23})\rb\frac{\,\lp \D_{D0} +\,\D_{D0} - \,2\,s_{23} + 2\,M^2_{D0} +
 M^2_{\pi} - l^2 \rp}{\D_{D^{+*}}\D_{D^0} \,\D_{\bar{D^0}}\;\D_a} \,.
\label{A1.2}
\eea
 The calculation of the integral in Eq. (\ref{A1.2})  follows the technique proposed in 
\cite{patWeak}. We got a similar amplitude  for the $K\pi$ channel and  the total decay amplitude,
$$\mathcal{A}_{B_c}(s_{23},s_{13})\,=\, A^{KK}_{B_c}(s_{23})+A^{K\pi}_{B_c}(s_{13})\, ,$$
becomes a function of both invariant masses, and provides a characteristic interference pattern in the Dalitz plot, 
as will be shown.
  
 \begin{figure}[ht!]
\begin{center}
\includegraphics[width=.45\columnwidth,angle=0]{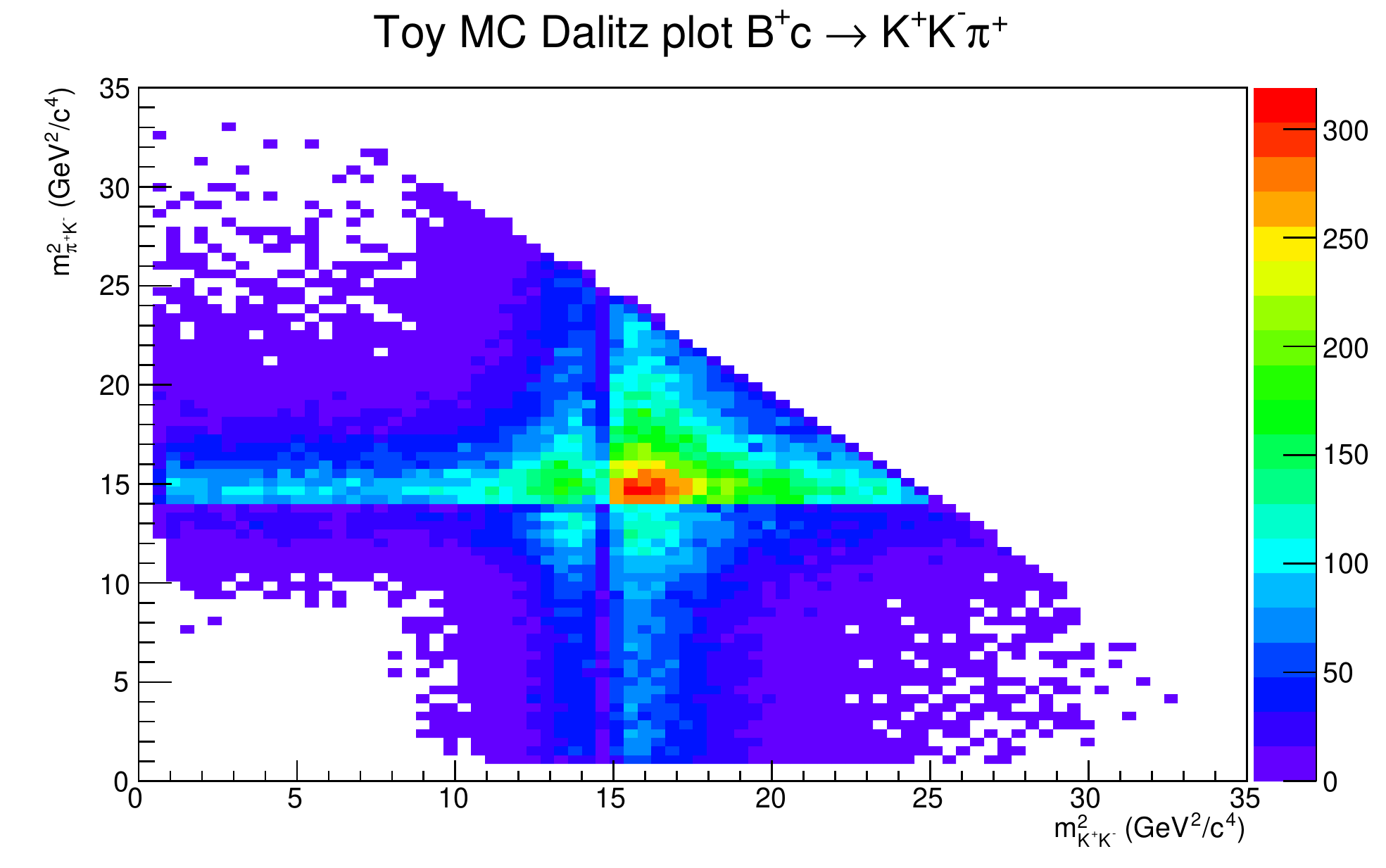}
\includegraphics[width=.45\columnwidth,angle=0]{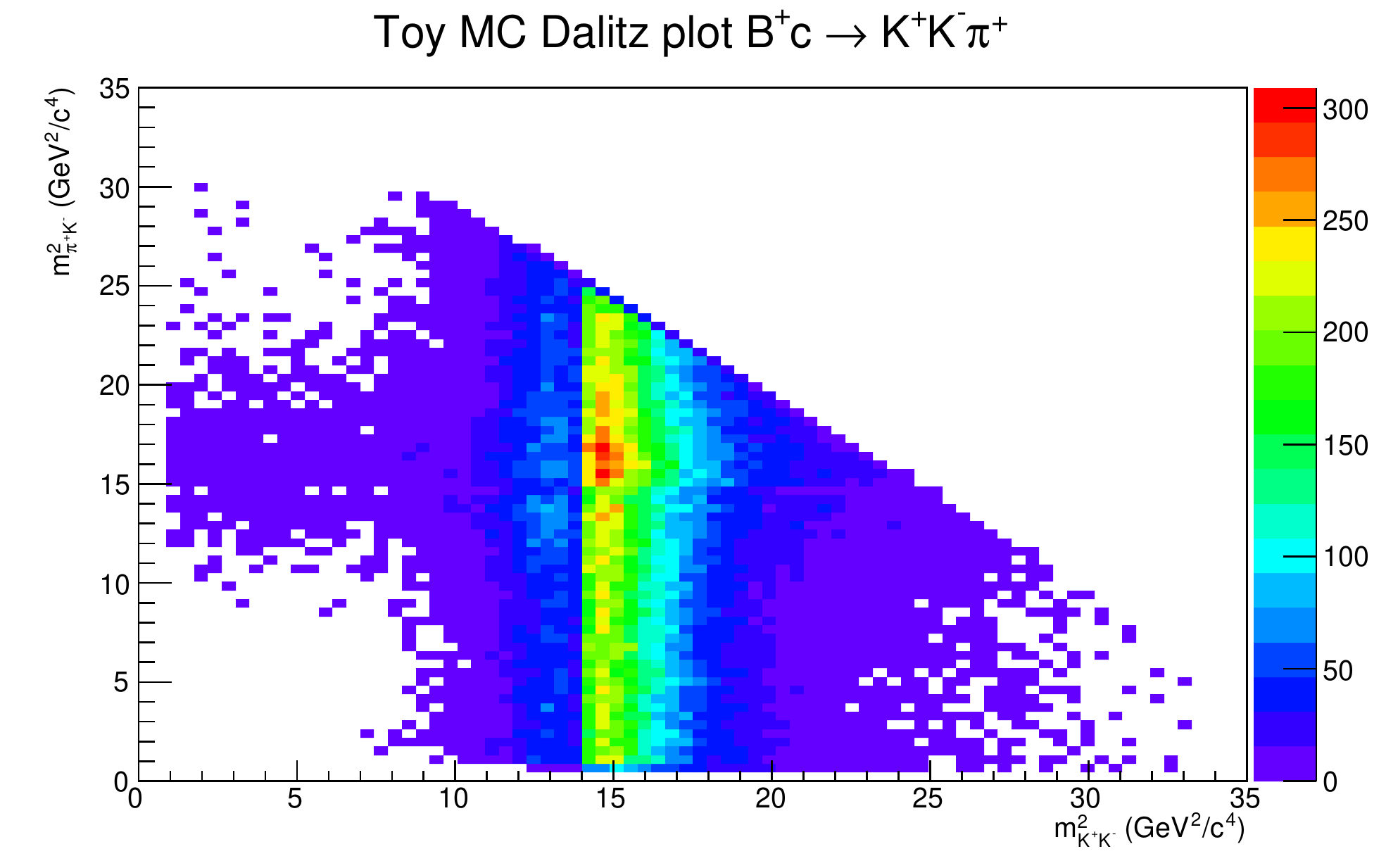}
\caption{ Dalitz plot for $\bckkp$ differential decay rates mediated by $D^{0} \bar{D^0}\to K^+K^-$  and  
$D^{-} \bar{D^+}_s\to K^-\pi^+$ . Left frame: same normalization for  $A^{K\pi}_{B_c}(s_{13})$ and $A^{KK}_{B_c}(s_{23})$ .
Right frame:  $A^{K\pi}_{B_c}(s_{13})$ normalized to 20\% of $A^{KK}_{B_c}(s_{23})$ (see text).  }
\end{center}
\label{fig5} 
\end{figure}
 
\section{Discussion}

\begin{figure}[ht!]
\begin{center}
\includegraphics[width=.35\columnwidth,angle=0]{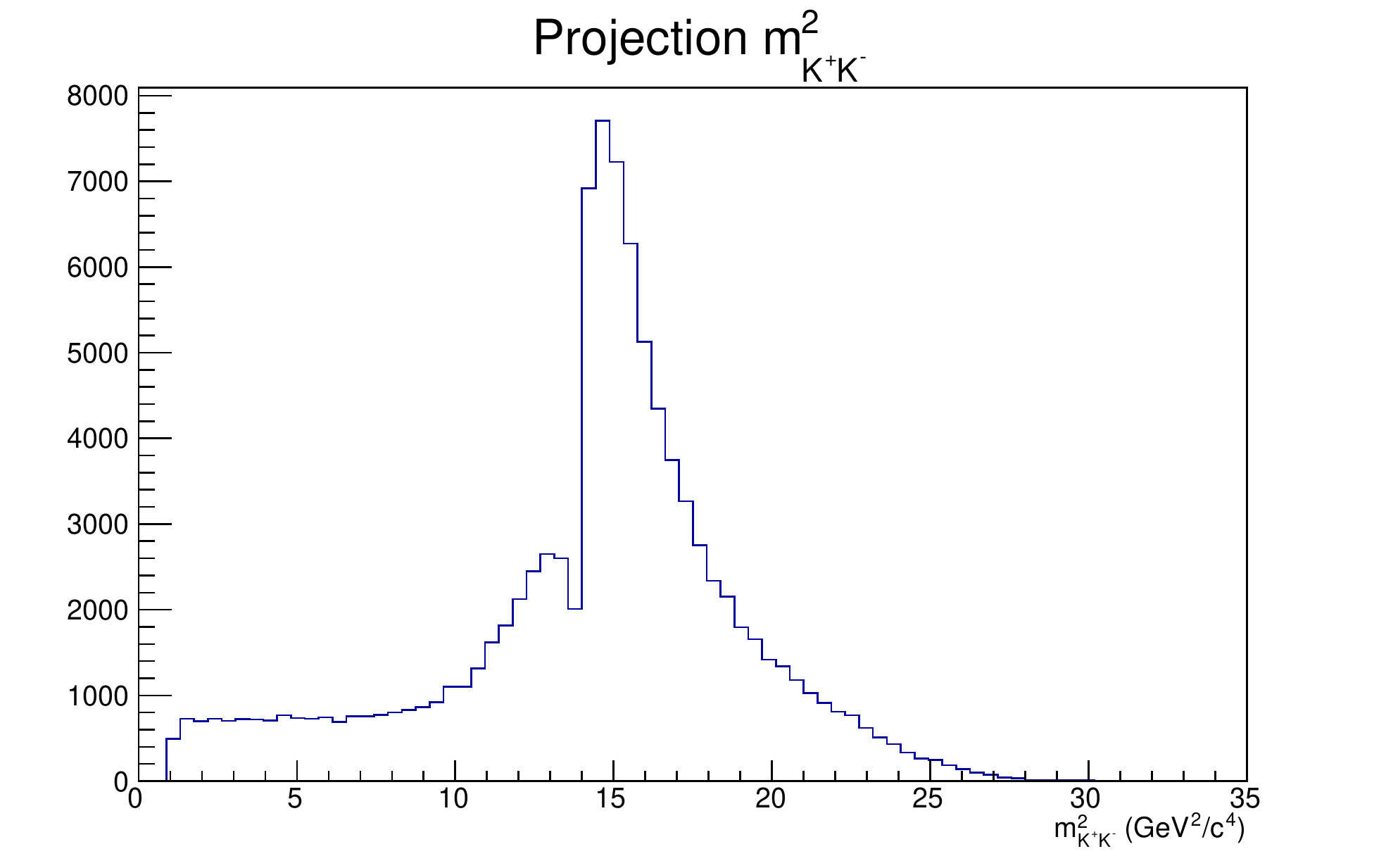}
\includegraphics[width=.35\columnwidth,angle=0]{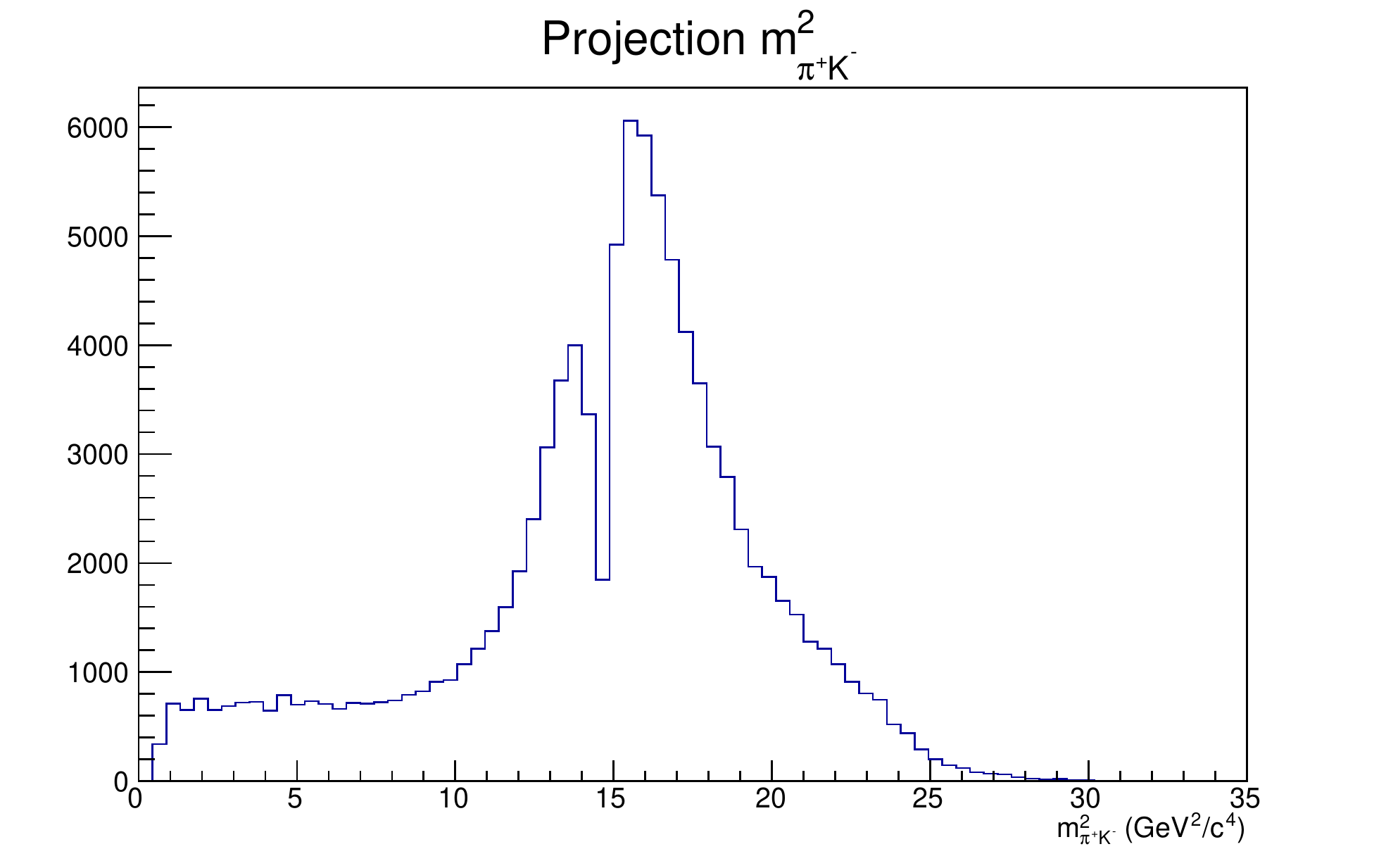}
\caption{ $\bckkp$ decay rate projections in the final states 
channels $ K^+K^- (s_{23})$ (left panel)  and $K^-\pi^+ (s_{13})$ (right panel).}
\end{center}
\label{fig6}
\end{figure}

The total decay amplitude  of the process $\bckkp$, $\mathcal{A}_{B_c}(s_{23},s_{13})$, 
was simulated in a Monte Carlo generator
 with 10,000 events using LAURA++  software \cite{laura}.
The resulting Dalitz plot is shown in Fig.~\ref{fig5} for two different  normalizations between the channels, 
whereas its projection in the different channels are shown 
in Fig. \ref{fig6}.

In the phenomenological scattering amplitudes $D^{0}\bar{D^{0}}\to K^-K^+ $ and  $D^{+}D_s^{-}\to\pi^+ K^-$, except 
for the thresholds ($s_{th\,PP}$) all the parameters in Eq. (\ref{t12c}) should be fixed by a fit to the data. 
In our toy amplitude the parameters were chosen based on our previous study~ \cite{us} and given in Table \ref{tab:1}.
 
 The  interference pattern presented in the Dalitz plot of Fig.~\ref{fig5} is a result of the sum of the 
 two amplitudes represented diagrammatically in Fig. 2. We assume the same branching fraction for the two charm decay channels,  that
 leads to the interference pattern  shown in the left frame of the figure. However,   in principle they 
 could have different production rates favouring one channel with respect
 to the other  one, which would change the Dalitz plot side bands but not the 
 minimum position associated with the double charm thresholds. This is illustrated in the right frame 
 of the figure, where we arbitrarily normalize $A^{K\pi}_{B_c}(s_{13})$ to 20\% of $A^{KK}_{B_c}(s_{23})$.

The projections of the differential decay rate in  the $KK$ and $K\pi$ channels are given respectively by the integration on the crossed-channels:
\begin{equation}
\int ds_{K\pi} |\mathcal{A}_{B_c}(s_{KK},s_{K\pi})|^2\,\,\, \text{and}\,\,\,\int ds_{KK} |\mathcal{A}_{B_c}(s_{KK},s_{K\pi})|^2\, ,
\end{equation}
and presented in Fig.~\ref{fig6} in the left and right frames. For this illustration we use
equal normalizations of   $A^{KK}_{B_c}(s_{23})$ and $A^{K\pi}_{B_c}(s_{13})$. In the figure
 one can see that the minima between the two bumps are in different 
positions in the left and right frames. The position of these minima  correspond to the opening of distinct
double charm final states, that have different thresholds and therefore they are displaced from one another. It is worthwhile to mention
that the unequal mass parameters  in the triangle and double charm scattering amplitudes affect the height of the
peaks at the right and left of the minima. A detailed study of these asymmetries can shed light on the parametrization of the
 transition amplitudes between the different two-body  channels. 

In the real data, the amplitudes of Fig. 2 should also interfere with other sources of 
non-resonant and resonant interactions.  This will not change the characteristic signature of the minimum 
found between the two bumps,  which are associated with the opening of  different double charm thresholds 
as shown in Fig. 4. 
If the $KK\pi$ final state is produced directly in the primary vertex, illustrated by
the diagram a) of Fig. 1 for the suppressed annihilation topology,  
one should expect, in analogy of what we know from the charmless B decays, that the 
final state interaction between the light mesons, will produce resonances in the low mass region 
in both $K^+K^-$ and $K^-\pi^+$ channels. 
However this is not what one could see
 from the actual data \cite{adlene} consistent with the present proposal.
 
In summary, our study shows that  the hadronic charm triangle amplitude with rescattering can play a  role in $\bckkp$ decay, producing a 
non-resonant process that leads to a signature in the middle of the Dalitz plot with 
a minimum close to $D\bar{D}$ threshold. 
We expect that the new data from LHCb run II, will be able to confirm  if the dominant mechanism to produce $K^+K^+\pi^-$ 
 through charm rescattering in  the  $B^+_c$  decay is supported by experiment.

\section*{ACKNOWLEDGMENTS}
This work was supported by Conselho Nacional de Desenvolvimento Cient\'{i}fico e Tecnol\'{o}gico (CNPq).  
 TF acknowledge the support from
Funda\c c\~ao de Amparo \`a Pesquisa do Estado de S\~ao Paulo (FAPESP grant \# 17/05660-0) and
  Project INCT-FNA Proc. No. 464898/2014-5. PCM acknowledge the support from TUM Physics faculty award for the promotion of gender equality in science.

\end{document}